\documentclass[twocolumn,showpacs,aps,nofootinbib,superscriptaddress]{revtex4-1}
\usepackage{color}
\RequirePackage{graphicx}
\begin{document}

\title{Radiative corrections of $O(\alpha)$ to $B^- \rightarrow V^0 \ell^- \bar{\nu}_{\ell}$ decays
}


\author{S. L. Tostado}
\email{stostado@fis.cinvestav.mx}
\affiliation{Departamento de F\'isica, Centro de Investigaci\'on y de Estudios Avanzados del Instituto Polit\'ecnico Nacional, 
Apdo. Postal 14-740, 07000 M\'exico D.F., M\'exico}
\author{G. L\'opez Castro}
\email{glopez@fis.cinvestav.mx}
\affiliation{Departamento de F\'isica, Centro de Investigaci\'on y de Estudios Avanzados del Instituto Polit\'ecnico Nacional,  Apdo. Postal 14-740, 07000 M\'exico D.F., M\'exico}
\affiliation{Instituto de F\'\i sica Corpuscular, CSIC- Universitat de Val\`encia, Apt. Correus 22085, E-46071 Val\`encia, Spain}





\begin{abstract}
The $O(\alpha)$ electromagnetic radiative corrections to the $B^- \rightarrow V^0 \ell^- \bar{\nu}_{\ell}$ ($V$ is a vector meson and $\ell$ a charged lepton) decay rates are evaluated using the cutoff method to regularize virtual corrections and incorporating intermediate resonance states in the real-photon amplitude to extend the region of validity of the soft-photon approximation. The electromagnetic and weak form factors of hadrons are assumed to vary smoothly over the energies of virtual and real photons under consideration.  The cutoff dependence of radiative corrections upon the scale $\Lambda$ that separates the long- and short-distance regimes is found to be mild and is considered as an uncertainty of the calculation.  Owing to partial cancellations of electromagnetic corrections evaluated over the three- and four-body regions of phase space, the photon-inclusive corrected rates are found to be dominated by the short-distance contribution. These corrections will be relevant for a precise determination of the $b$ quark mixing angles by testing isospin symmetry when measurements of semileptonic rates of charged and neutral $B$ mesons at the few percent level become available.  For completeness, we also provide numerical values of radiative corrections in the three-body region of the Dalitz plot distributions of these decays. 
\end{abstract}

\maketitle

\section{Introduction}
\label{intro}
A precise determination of the $|V_{ub}|$ quark mixing at a few of percent level, is crucial for future tests of the Standard Model (SM) picture of CP violation \cite{ckm} as it fixes one of the sides of the $db$ unitarity triangle. A combination of inclusive and exclusive channels of charmless leptonic and semileptonic decays of $b$ hadrons can serve this purpose in two ways.  On the one hand, they can provide a consistency check of the models employed to extract $|V_{ub}|$; this in turn should yield reduced uncertainties by using the average of independent  and consistent determinations.  As an example, let us remind that the consistency check of the set of  data for superallowed Fermi transitions (kaon semileptonic decays), which require a detailed calculation of all sources of isospin breaking corrections, allows a determination of $|V_{ud}|$ ($|V_{us}|$) at the per-mille (one percent) level \cite{pdg2014}.  Given that different charmless semileptonic decays of charged and neutral $B\to (P,V)$ transitions are related by isospin symmetry, similar consistency checks can be attempted with future measurements in order to reach a better accuracy in the extraction of $|V_{ub}|$.

 Several leptonic and semileptonic exclusive decays induced by the $b \to u\ell \nu_{\ell}$ transition can be used to extract $|V_{ub}|$ in an independent way.  The cleanest channel owing to the better control on theoretical and experimental inputs is $B\to \pi\ell \nu_{\ell}$, yielding $|V_{ub}|_{\rm excl}=(3.28\pm 0.29)\times 10^{-3}$ \cite{pdg2014}. This relies on the most precise measurements of the total and differential rates \cite{btopi-cleo,btopi-belle,btopi-babar}, as well as on the lattice \cite{lattice} and the Light-cone QCD  sum rules (LCSR) \cite{lcsr} calculations\footnote{New LCSR calculations of the $B\to \pi$ form factors may actually further decrease the extracted value of  $|V_{ub}|$ by up to 10\% \cite{lcsr1}.} of the  form factor. When compared to the most precise determination from inclusive $b\to u \ell \nu$ transitions one is led to discrepant results  $\Delta |V_{ub}|=|V_{ub}|_{\rm incl}-|V_{ub}|_{\rm excl}=(1.13\pm 0.36)\times 10^{-3}$ \cite{pdg2014}, with errors added in quadrature. Improvement in both, experiment and theory, will play a crucial role in understanding and solving this  discrepancy\footnote{ An interesting suggestion that duality violations may occur close to thresholds and would lead to smaller values of $|V_{ub}|_{\rm incl}$, has been put forward recently \cite{bigi2015}.}. On the other hand, as long as new measurements of $B^{+}\to \tau^+\nu_{\tau}$ \cite{btotaunuexp} are getting in better agreement with the SM prediction \cite{btotaunuteo}, it will become a useful and independent test in the determination of $|V_{ub}|$. Along this goal, a recent measurement of the ratio $(\Lambda_b\to \Lambda)/(\Lambda_b\to \Lambda_c)$ of partially integrated semileptonic rates by the LHCb collaboration \cite{vubnature}, combined with the exclusive value of $|V_{cb}|$ \cite{pdg2014} and Lattice QCD calculations of the relevant hadronic form factors \cite{baryon-lat}, have provided the first competitive determination of $|V_{ub}|$ from baryon semileptonic decays: $|V_{ub}|_{\Lambda_b}=(3.27\pm 0.15_{\rm exp}\pm 0.16_{\rm Latt}\pm 0.06_{V_{cb}})\times 10^{-3} $ \cite{vubnature}.

In this work we are concerned with the calculation of electromagnetic radiative corrections to  $B^{\pm}\to V^0$ semileptonic transitions, a useful input in testing isospin symmetry in those $B$ decays. In the isospin symmetry limit, the following relations hold\footnote{Similar relations hold by replacing $\rho \to \pi$ mesons.}
\begin{eqnarray}
\Gamma^I(B^0\to \rho^-\ell^+\nu_{\ell})&=&2\Gamma^I(B^+\to \rho^0\ell^+\nu_{\ell})\nonumber \\ 
&=&2\Gamma^I(B^+\to \omega\ell^+\nu_{\ell})\ , \label{isospin}
\end{eqnarray}
which seem to be supported by experimental data within current experimental errors (see Table \ref{table1}). Here we will assume the narrow width approximation for vector mesons; in practice, vector mesons are unstable particles that are reconstructed from the invariant-mass distribution of suitable decay channels \cite{quasistable1,quasistable2}.  For instance,  $B^{\pm}\to \rho^0\ell^{\pm}\nu_{\ell}$ must be extracted by choosing a narrow window in the $2\pi$ invariant-mass distribution of the full  $B^{\pm}\to \pi^+ \pi^-\ell^{\pm}\nu_{\ell}$ process. Radiative corrections to this decay requires the consideration of additional electromagnetic interactions involving final state particles, which are not present in decays with neutral vector mesons. This and the modelling of the hadronic current increases considerably the difficulty of calculations. Therefore, considering vector mesons as asymptotic states is an intrinsic limitation of our present calculation.
\begin{table}[h]
\caption{\footnotesize Rates ($\times 10^{8}$ s$^{-1}$) for $B\to V$ charmless semileptonic transitions using the average of experimental branching fractions  measured by BABAR \cite{br-babar}, BELLE \cite{br-belle} and CLEO \cite{br-cleo}, and the lifetimes \cite{pdg2014} of $B$ mesons as inputs.}
\begin{tabular}{|c|c|} 
\hline
Channel & PDG 2014 \cite{pdg2014} \\
\hline
$B^0\to \rho^-\ell^+\nu_{\ell}$ & 1.61$\pm$ 0.21 \\
$B^+\to \rho^0\ell^+\nu_{\ell}$ &0.87$\pm$ 0.14\\
$B^+\to \omega\ell^+\nu_{\ell}$ &0.73$\pm$ 0.06\\
\hline
\end{tabular}\label{table1}
\end{table}

Departures from the isospin symmetry relations (\ref{isospin}) are expected at the few percent level owing to effects of electromagnetism and $u-d$ quark mass difference. The observable effects of isospin breaking  manifest  in hadron masses, hadronic form factors \cite{quasistable1,rho-omega} and long-distance electromagnetic radiative corrections. Future improved measurements of semileptonic $B\to (\pi, \rho, \omega)$ transitions will require consideration of isospin breaking effects. 

In this paper we contribute to this goal with  the calculation of long-distance radiative corrections to $B^- \to (\rho^0,\omega)\ell^- \nu_{\ell}$ decays, which to the best of our knowledge have not been considered before. Our results can be applied  to $B^- \to D^{*0}\ell^-\nu_{\ell}$ decays as well, within the validity of the underlying assumptions. We compute both, the long-distance corrections to the photon-inclusive decay rates and the three-body region of the Dalitz plot distribution of these decays. These electromagnetic corrections should affect the equality between the first two decay rates in Eq. (\ref{isospin}), but not the ratio between the last two. Instead, $\rho^0-\omega$ mixing, which originates mainly from the $u-d$ quark mass difference, affects only the decay rates with $(\rho^0,\omega)$ mesons in the final state \cite{rho-omega}.

Electromagnetic corrections for $B$ meson decays are more complicated than in light hadrons due to the presence of hard-photons. Here we deal with this problem by cutting photon momenta in virtual corrections at some relatively small cutoff scale where photon-meson interactions can be loosely approximated by scalar QED. For real photon emission, we include the effects of resonance contributions in order to extend the validity of the soft-photon approximation. This sets a limitation of our calculation, which may weakens the validity of the approach for charmless semileptonic transitions, but can work better for the charmfull case.

\section{Radiative corrections to $B^-\to V^0$ semileptonic transitions}

There are only a few works related to the calculation of radiative corrections to semileptonic decays of $B$ mesons \cite{atwood-marciano,bernlochner,Cirigliano2006}. Among the main reasons one finds the experimental difficulty to reach the few percent level accuracy in the measurements of the branching fractions and the theoretical limitation imposed by our knowledge of the meson-photon interactions in  $B$ meson decays. A first calculation in the case of $B\to P\ell \nu_{\ell}$ transitions was made by incorporating the resummation of real soft-photon emission and virtual corrections \cite{atwood-marciano}. Other works \cite{bernlochner} follow an approach similar to the ones adopted in analogous kaon semileptonic decays \cite{Ginsberg,Cirigliano,Andre,Garcia,ruben1}. An important limitation in the  attempts to compute those radiative corrections has to do with the poor knowledge of the photon-meson interactions at wavelength in the transition between the long- and short-distance regimes. Thus, most of these calculations, as well as the present one, suffer from  limitations related to the used approximations which may be well justified for light mesons and baryons but not necessarily for $B$ mesons.

The $O(\alpha)$ electromagnetic radiative corrections involve the emission and/or re-absorption of virtual and real photons, with virtual photon energies ranging from zero to infinity. By long-distance (LD) corrections we mean, emission/absorption of  real and virtual photons with small momenta $k$  ($ <\Lambda_1$)  such that they cannot resolve the structure of hadrons; at these low photon momenta, the point-like approximation for hadrons and the use of scalar QED for describing photon-meson interactions should be a good approximation. For photon momenta larger than a scale  $\Lambda_2$, photons resolve the charged components of mesons and the description in terms of  the electroweak (EW) theory to describe short-distance (SD) interactions is justified. 

The transition region $\Lambda_1 (\rm {a\ few- hundreds\ MeV's}) < k < \Lambda_2 ({\rm a \ few\ GeV})$ between the short- and long-distance approximations is more difficult to evaluate and a description based on exchange of meson resonances may be an appropriate model. To the best of our knowledge, an extension of the long-distance approximation to cover photon energies around 1 GeV has been done only in Ref \cite{Decker} by including resonance degrees of freedom in $\tau \to M \nu(\gamma)$ and $M\to \ell \bar{\nu}_{\ell} (\gamma)$ decays. Ref \cite{Decker} has found that the long-distance corrections to the ratio $R_{\tau/\pi}\equiv \Gamma(\tau \to \pi \nu)/\Gamma(\pi \to \mu \nu)$ computed in the scalar QED approximation using a scale $\mu_{\rm cut}$ to cut the integration over virtual photons differs from the improved calculation that include resonances by less than 0.4\%. Also, it was found \cite{Decker} that the dependence upon the scale $\mu_{\rm cut}$ (assumed to separate short- and long-distance regions) largely cancel in the ratio $R_{\tau/\pi}$ and, actually, it becomes insignificant for $\mu_{\rm cut}> 1.5$ GeV's. However, the radiative corrections to the individual decay rates keeps a substantial dependence  upon $\mu_{\rm cut}$ \cite{Decker}. 

  In practice a single scale $\Lambda$ (which plays the same role as $\mu_{\rm cut}$) can be used to separate the long- and short-distance regimes, and further assumptions about a small variation of hadronic structure have to be adopted. Thus, in the absence of a widely accepted prescription that perfectly matches the corrections in these two regions, an unavoidable dependence of radiative corrections  $\delta_{ T}$ upon the cutoff scale $\Lambda$ naturally arises \footnote{Note however that in the case of the ChPT calculation of radiative corrections to $K_{\ell 3}$ decays a convenient choice of the relevant local counterterms can serve to cancel the cutoff scale dependence in $\delta_T$ \cite{Cirigliano}.}. As long as the radiative correction $\delta_{ T}=\delta_{\rm LD}(\Lambda)+\delta_{\rm SD}(\Lambda)$ does not strongly depend on the cutoff scale, we may be more confident upon the model assumptions used to compute the long-distance corrections. In the present paper we will adhere to this procedure in the calculation of the integrated rates and assign an uncertainty due to the $\Lambda$-dependence owing to un-matching of long- and short-distance radiative corrections. Since our results will be suitable for photon inclusive rates, in order to extend the validity of real-photon corrections to higher photon momenta, we also consider some of their model-dependent terms. Certainly, this and the difficulty to assess the uncertainty of other possible structure-dependent contributions is one of the limitations of our approach.

The integrated rate of $B^- \to V^0 \ell^-\nu_{\ell}$ decays ($V=\rho, \omega, D^*$) including corrections of $O(\alpha)$ will be written as
\begin{eqnarray}\label{factorize}
\Gamma&=&\Gamma^0+\Gamma^1_{III}+\Gamma^1_{IV} \nonumber \\
&=& \Gamma^0( 1+\delta^1_{\rm LD})(1+\delta^1_{\rm SD})\ . 
\end{eqnarray}
In the first line of (\ref{factorize}), the superscripts in the rates denotes the order in $\alpha$ and the subscripts the region of phase-space corresponding to three- and four-body kinematics (the four-body region is accessible only when real-photons are emitted). Up to $O(\alpha)$, the LD and SD corrections (superscripts denote the order in $\alpha$) can be factorized according to the second line in Eq. (\ref{factorize}).  

The SD corrections to semileptonic decays are finite in the electroweak theory \cite{sirlin1978,ms1988}. In the dominant logarithmic approximation  the $O(\alpha)$ corrections depend upon the separation scale $\Lambda$ as follows ($m_Z$ is the $Z$ boson mass) 
\begin{equation}\label{sdcorrection}
\delta^1_{\rm SD}=\frac{2\alpha}{\pi}\ln\left(\frac{m_Z}{\Lambda}\right)\ .
\end{equation}
 It is customary to choose the lower cutoff of SD corrections as the mass of the decaying particle; however, we will keep it explicitly to study the cutoff scale dependence of the full radiative corrections. The large  EW logarithms can be summed to all orders using the renormalization group \cite{ms1986}; the dominant part of $O(\alpha \alpha_s)$ corrections have also been calculated \cite{sirlin1978} and resummed to all orders \cite{erler-rmf}. For consistency, here we keep only the leading approximation (\ref{sdcorrection}) in our evaluation.

In order to compute the long-distance part of the radiative corrections, let us consider for definiteness the decay $B^{-}(P, M)\rightarrow V^{0}(P_V, m_V) \ell^{-}(p,m) \bar{\nu}_{\ell}(p',0)$, where $V$ is a vector meson; the first character within parentheses denote the four-momentum of each particle and the second its corresponding mass. At the tree-level, the decay amplitude is:
\begin{equation}\label{treelevel}
{\cal M}^0 = \frac{G_F}{\sqrt{2}}  V_{qb} c_V W_{\nu}(P_V,P)L^{\nu},
\end{equation}
where $G_F$ is the Fermi constant, $V_{qb}$ ($q=u,c$) the relevant CKM matrix element,  $L^{\nu} = \bar{u}_\ell \gamma^{\nu} (1-\gamma_5) v_{\nu} $ represents the leptonic current, and $c_V=1/\sqrt{2}$ is the Clebsh-Gordan coefficient for $V=(\rho^0, \omega)$ mesons. The hadronic matrix element can be parametrized in terms of four, $q^2=(P-P_V)^2$-dependent form factors\footnote{With $A = A_{0} - \frac{M+m_V}{2 m_V} A_{1} + \frac{M-m_V}{2 m_V} A_{2}$.}, which we choose as $(V,A_{1},A_{2},A)$
\begin{eqnarray}\label{hme}
W_{\nu}(P_V,P) &=&  \frac{2V^{B\to V}}{M + m_{V}} \epsilon_{\nu \alpha \beta \gamma }\varphi^{\alpha} P^{\beta}P_{V}^{\gamma}   \nonumber \\
&& \!\!\!\!\!\!\!\!\!\!\!\!\!\!\!\!\!\!\! -i (M+m_{V}) A^{B\to V}_{1} \varphi_{\nu}  +i \frac{ A^{B\to V}_{2}}{M + m_{V}} q \cdot \varphi (P+ P_{V})_{\nu } \nonumber \\
&& -i \frac{2 m_{V} A^{B\to V}}{q^{2}}  q \cdot \varphi q_{\nu } \ .
\end{eqnarray} 
The four-vector $\varphi$ denotes the vector meson polarization, with the orthogonality condition $P_V\cdot \varphi=0$. The form factors are specific to the $B\to V$ transition, although hereof we will disregard their superscripts. At higher orders, the factorization of the amplitude is not valid and the effective form factors may depend upon another invariant variable, for example $u=(P-p)^2$.

\subsection{Virtual-photon corrections}

\begin{figure}\centering
\includegraphics[scale=0.5]{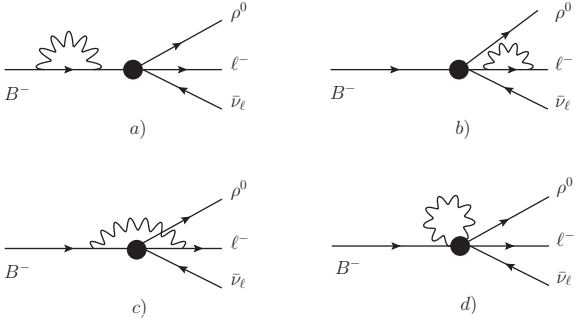}
\caption{\footnotesize QED virtual corrections to $B^{-}\rightarrow \rho^{0} \ell^{-} \bar{\nu}_{\ell}$. Figs. $a)$ and $b)$ correspond to the self-energies of charged particles and $c),\ d)$ to vertex contributions. }
\label{diagvirtual}
\end{figure}

The virtual QED corrections of $O(\alpha)$ are shown in Fig. \ref{diagvirtual}. We will consider that the momenta of virtual photons are not very large (typically $< 1.5$ GeV) so that scalar QED can be used for the electromagnetic vertices of the $B^-$ meson and the weak hadronic vertex does not vary strongly.

The charged meson ($\delta Z_B$) and lepton ($\delta Z_{\ell}$)  self-energies (Fig. \ref{diagvirtual} $a$ and $b$) contribute to the wave function renormalization of the charged particles. In terms of  ${\cal M}^{0}$ their contributions to the decay amplitude are
\begin{equation}\label{wfr} 
{\cal M}^1_{a+b} = {\cal M}^{0}  \frac{1}{2}\left( \delta Z_{\ell} + \delta Z_{B} \right),
\end{equation}
where 
\begin{eqnarray}
\delta Z_{\ell} &=& \frac{\textstyle \alpha}{\textstyle 4 \pi} \left(\frac{}{} 2 - {\rm B}_0[m^{2},0,m^{2}] +4m^{2}{\rm B}'_0[m^{2},\lambda^{2},m^{2}] \right)  \nonumber \\ 
\delta Z_{B} &=& \frac{\textstyle \alpha}{\textstyle 4 \pi} \left( \frac{}{} 2{\rm B}^{M}_{0}[M^{2},0,M^{2}] +4M^{2}{\rm B}'^{M}_{0}[M^{2},\lambda^{2},M^{2}] \right)  \nonumber \ . 
\end{eqnarray}
Here, B$^{(M)}_0[\cdots]$ and B$'^{(M)}_{0}[\cdots]$ are Passarino-Veltman functions corresponding to the scalar two-point integral and its derivative \cite{Passarino}. Because of the infrared (IR) singularity in B$'^{(M)}_{0}$, we have provided a fictitious mass $\lambda$ to the photon. The ultraviolet (UV) singularity contained in B$_{0}$ can be regulated by using the cutoff  $\Lambda$ as the maximum scale at which the long-distance approximation is expected to be valid; the expressions for the Passarino-Veltman functions required in this work are shown in  appendix A. 

In order to treat the remaining corrections shown in Figure \ref{diagvirtual}c,d we follow Sirlin's prescription \cite{Sirlin} to separate their model-independent (basically scalar QED) and model-dependent contributions. This leaves us with the amplitude
\begin{eqnarray}\label{vertexcor}
{\cal M}^1_{c+d} &=& -e^2 \frac{G_F}{\sqrt{2}} \bar{u}_{\ell} \int \frac{{\rm d}^4 k}{(2\pi^4)} D^{\mu \nu}(k) \Delta (P+k) \nonumber \\
&& \times \left[(2P +k)_{\mu}W_{\lambda}(P_V,P) + \frac{}{}T_{\mu\lambda}(P,P_V,k)\right] \nonumber \\
&& \times \gamma_{\nu} S_F(p+k) \gamma^{\lambda}(1-\gamma_5)v_{\nu_{\ell}} ,
\end{eqnarray}
where $D^{\mu \nu}(k), \Delta (P+k)$ and $S_F(p+k)$ correspond to the propagators of the photon (chosen in this work in the Feynman gauge), the pseudoscalar meson and the lepton, respectively. Thus, the  model-independent part of the virtual correction corresponding to the photon exchange between the initial meson and the charged lepton (Fig. \ref{diagvirtual} c) gives rise to the first term in square brackets of Eq. (\ref{vertexcor}). The model-dependent part is encoded in $ T_{\mu\lambda}$, where gauge invariance requires $k^{\mu} T_{\mu\lambda}=0$, which can be verified by applying the generalized Ward identity. This model-dependent contribution can be included in a re-definition of the hadronic form factors as done for instance in \cite{Garcia,ruben1,Sirlin} or evaluated explicitly as in Ref. \cite{Decker} in a given model. 

For the model-independent part of the virtual corrections we have (remember $u=\!(P-p)^2$)
\begin{eqnarray}\label{vertex}
{\cal M}^1_{\rm c+d, MI}&=&\frac{\alpha}{4\pi}{\cal M}^{0}\left\lbrace \frac{}{} {\rm B}_{0}^M[M^{2},0,M^{2}] \right. \nonumber \\ 
&& \left.+ 4 p\cdot P {\rm C}_{0}[m^2,M^2,u,m^2,\lambda^{2},M^{2}]  \right. \nonumber \\ 
&& \left. +2 M^2 {\rm C}_{1}[m^2,M^2,u,m^2,0,M^{2}]  \right. \nonumber \\ 
&& \left.+ 4 P\cdot p {\rm C}_{2}[m^2,M^2,u,m^2,0,M^{2}] \frac{}{} \right\rbrace \nonumber \\
&& -\frac{\alpha}{4\pi} {\cal M}_{\rm NF}{\rm C}_{2}[m^2,M^2,u,m^2,0,M^{2}] \ ,
\end{eqnarray} 
where C$_{i}[m^2,M^2,u,m^2,\lambda^{2},M^{2}]$ ($i=0,1,2$) denote the three point functions.  We have used again $\lambda$ to isolate the IR-divergence contained in C$_0$, while C$_{1,2}$ can be evaluated in the limit $\lambda\rightarrow 0$, and be written in terms of $\Lambda$-dependent scalar two point functions as is shown in the Appendix A. The last term in Eq.(\ref{vertex}), where we have defined ${\cal M}_{\rm NF}=\frac{G_F}{\sqrt{2}} m V_{ub} W^{\mu}(P_V,P) \bar{u}_\ell \not{P}\gamma_{\mu}(1-\gamma_5)v_{\bar{\nu}_\ell}$, corresponds  to a non-factorizable (NF)  amplitude, it is IR safe and gives a negligible contribution for light charged leptons in the final state.

The regulated infrared divergences can be isolated and written explicitly, while the UV divergences are implicit in the Passarino-Veltman functions. By collecting the results of the virtual corrections, Eqs. (\ref{wfr})-(\ref{vertex}), we can write the radiatively corrected rate as follows: 
\begin{eqnarray}\label{wvirtual}
\frac{{\rm d}^2{\Gamma}^1_{v}}{{\rm d}E{\rm d}E_V} &=& \frac{\alpha}{4\pi} \frac{{\rm d}{\Gamma}^{0}}{{\rm d}E{\rm d}E_V} \left\lbrace -6 + 2 {\rm ln}\left(\frac{M^2}{\lambda^2}\right)+ 2{\ln}\left(\frac{m^2}{\lambda^2}\right) \right. \nonumber \\ && \left. + 2\left(\frac{3}{2} +\frac{1-\beta^2}{\beta^2} \right) {\rm B}_0[m^{2},0,m^{2}] \right. \nonumber \\
&& \left. + 2\left(2 - \frac{M}{E\beta^2} \right) {\rm B}^{M}_{0}[M^{2},0,M^{2}] \right. \nonumber \\ 
&& -2\left(2 +\frac{1-\beta^2}{\beta^2} - \frac{M}{E\beta^2} \right){\rm B}^{lM}_{0}[u,m^2,M^{2}] \nonumber \\ && 
 -4 M E\ {\rm F}_{2}(E) \!\! \left. -\frac{2}{\beta}\left[{\ln}\left(\frac{1+\beta}{1-\beta}\right) {\ln}\left(\frac{u}{\lambda^2}\right) \right]  \right\rbrace \nonumber \\ 
&& -\frac{\alpha}{4\pi} \frac{{\rm d}{\Gamma}^{1}_{\rm NF}}{{\rm d}E{\rm d}E_V},
\end{eqnarray}
where $\beta (E)$ is the velocity (energy) of the charged lepton and $E_V$ the energy of the vector meson  in the rest frame of the decaying particle. 
The expression for the function ${\rm F}_{2}(E)$ can be found in the Appendix A. In the above result, d$\Gamma^0$/d$E$d$E_V$ corresponds to the differential decay rate of $B^-\to V^0\ell^-\bar{\nu}_{\ell}$ at the tree-level, while $d\Gamma^1_{NF}/dEdE_V$ arises from the interference of the tree-level amplitude and the last term in Eq. (\ref{vertex}).

\subsection{Real-photon corrections}

In this subsection we consider the real photon corrections. As is well known, the cancellation of IR divergences occurs when we add incoherently the decay rates of a process with virtual and real-photon corrections. We will first consider the soft-photon approximation (SPA) for the decay amplitude and assume scalar QED for the electromagnetic vertex of $B^-$ meson. This {\it model-independent} approximation may be questionable and in principle it would be more appropriate only for the end-point region of the lepton spectrum \cite{atwood-marciano}. Then, we will also include {\it model-dependent} terms of order zero and one in the photon momentum $k$. The general form of the decay amplitude can be written as ${\cal M}_{\gamma}= {\cal M}^{\rm Low}+ {\cal M}^{\rm MD}$, where ${\cal M}^{\rm Low}$ is the Low's soft-photon amplitude \cite{Low} which contains terms of order $k^{-1}$ and $k^0$, only; the second term ${\cal M}^{MD}$ is explicitly model-dependent and of $O(k)$. In the case of $K_{\ell 3}$ decays the model-dependent terms of $O(k^0)$ were found to be negligible\footnote{The expression of those terms depends on the specific model for the weak form factors;  in the case of $K_{\ell 3}$ decays their effects in the radiative corrections are at the one per mille level \cite{Andre}.}, however they deserve a proper study in $B$ decays.

\subsubsection{Model-independent corrections}

\begin{figure}\centering
\includegraphics[scale=0.5]{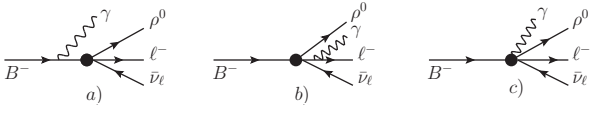}
\caption{\footnotesize Feynman diagrams for real photon emission. }
\label{diagbrem}
\end{figure}

By taking into account the contributions in Fig. \ref{diagbrem}  one gets the decay amplitude in the Low's approximation \cite{Low,Fearing}
\begin{eqnarray}\label{bremss}
{\cal M}^{\rm Low} &=& e \frac{G_{F}}{\sqrt{2}} V_{qb} c_V \left\lbrace \frac{}{} W_{\mu}(P_V,P) \right. \nonumber \\
&& ~ \!\! \times \bar{u}_{\ell} \left[-\frac{ P\cdot \varepsilon}{P \cdot k}+\frac{2p\cdot \varepsilon + \not{\varepsilon} \not{k}}{2 p\cdot k} \right]  
 \gamma^{\mu}(1-\gamma_5)v_{\nu_{\ell}} \nonumber \\
&& \!\!\! \left.  + L^{\mu}D^{\lambda} W_{\mu \lambda} - 2 P_V \cdot D ~L^{\mu} \frac{\partial W_{\mu}(P_V,P)}{\partial q^2} \right\rbrace ,
\end{eqnarray}
where $\varepsilon$  denotes the  polarization four-vector  of the real photon, with $k \cdot \varepsilon=0$. We have defined $D^{\lambda}= (\varepsilon \cdot P/k\cdot P)k^{\lambda} - \varepsilon^{\lambda}$,  $L$ and $W$ are the same as in Eq. (\ref{treelevel}), and
\begin{eqnarray}
W_{\mu \lambda} &=& -\frac{2}{M+m_V} V \epsilon_{\mu\lambda\nu\alpha} \varphi^{\nu}_V P_V^{\alpha} + i \frac{q\cdot \varphi}{M+m_V}A_2 \delta_{\mu \lambda}  \nonumber \\
&& ~ + i \frac{A_2}{M+m_V} (P+P_V)_{\mu}\varphi_{\lambda} -  2i \frac{m_V A}{q^2} q_{\mu} \varphi_{\lambda} \nonumber \\
&& - 2i \frac{m_V A}{q^2} q\cdot \varphi \delta_{\mu \lambda} ~.
\end{eqnarray}
As required by Low's theorem \cite{Low}, Eq. (\ref{bremss}) contains terms up to $O(k^0)$ which are fixed by gauge-invariance  and depend only upon the form factors of the non-radiative decay amplitude. 
The partial derivative in the last term of Eq. (\ref{bremss}) must be taken over explicit $q^2$-dependent form factors in $W$. For the purposes of numerical calculations, we will consider the evaluation of the last two terms in Eq. (\ref{bremss}) as a part of the model-dependent contribution (see next subsection).  In the remainder of this subsection, we will focus only on the first term of Eq. (\ref{bremss}) which we call the model-independent piece of real photon corrections.

After integrating over the momentum of the massless particles, we write the differential decay width corresponding to the model-independent term (MI)  as 
\begin{eqnarray}\label{bremss1}
\frac{{\rm d}^{2}{\Gamma}^{MI }}{{\rm d}E \ {\rm d}E_{V}}&=&\frac{\alpha}{32\pi M(2\pi)^3} \left[ |\overline{{\cal M}^0}|^2 \! I_{0}(E,E_V,\lambda^2)\right. \nonumber \\
&&\left. \!\!\!\!\!\!\!\!\!\!\!\!\!\!\!\!\!+ \frac{G_F^2}{2} |V_{ub}|^2 \!\! \int_{0}^{x_{max}}\!\!\!\!\! {\rm d}x \sum_{m,n} C_{m,n} I_{m,n}(E,E_V,x)   \right] 
\end{eqnarray}
where $|\overline{{\cal M}^0}|^2$ is the unpolarized squared amplitude at lowest order, and $(E,E_V)$ are the lepton and vector-meson energies in the rest frame of the decaying particle ($q^2=M^2+m_V^2-2ME_V,\ u=M^2+m^2-2ME$). Note however that, in radiative decays, the phase space region accessible to $(E,E_V)$ is larger than in the three-body decays and will depend upon the energies carried out by the massless particles through the invariant mass variable $x\equiv (q+k)^2$ \cite{Ginsberg}.

In a similar way as in Ref. \cite{Ginsberg},  the first term in Eq. (\ref{bremss1}) contains the  contribution proportional to the zeroth order decay width and carries all the infrared divergences of real-photon corrections, regulated as in the previous case by $\lambda$,  
\begin{eqnarray}\label{Izero}
 I_{0}(E,E_{V},\lambda^2) &=& \int_{\lambda^2}^{x_{max}}\!\!\!\! {\rm d}x \! \left[ 2 P\cdot p ~I_{1,1}(E,E_{V},x) \right. 
\nonumber \\ && \left. \!\!\!\!\!\!\!\!\!\!\!\!\!\!\!\!\!\! - M^2 I_{0,2}(E,E_{V},x) - m^2 I_{2,0}(E,E_{V},x) \right] \ .
\end{eqnarray}
Here, $x_{max}=(M-E-E_V)^2-|\mathbf{p_V}-\mathbf{p}|^2$ and the region of $(E,E_V)$ accessible in the four-body decay can be found in \cite{Ginsberg}.
The integrals ($P_f\equiv P_V+p+p'+k$)
\begin{equation}
I_{m,n}(E,E_V,x)\equiv \frac{1}{2\pi} \int \frac{\rm{d}^3 p'}{E'} \frac{\rm{d}^3 k}{k_0} \frac{\delta^{(4)}(P-P_f)}{(p\cdot k)^m (P\cdot k)^n}\ ,
\end{equation}
required for $K_{\ell 3}$ decays,  were first evaluated in \cite{Ginsberg} and verified in \cite{Cirigliano}. The coefficients  $C_{m,n}$ of the additional terms that appear in Eq. (\ref{bremss1})  can be found in the Appendix B. 
The second term in Eq. (\ref{bremss1}), is infrared finite and can be evaluated by setting the photon mass to zero. 

The region of integration of the variable $x$ is identical to the one considered in \cite{Ginsberg}, after tanking into account the corresponding changes in the hadron masses. The domain of the Dalitz plot ($E, E_{\rho}$) is shown in Figure \ref{dalitz} for the $B^-\to \rho^0\mu^-\nu_{\mu}$ channel (hereafter we will refer to this channel;  under the corresponding approximations, the results can be applied to channels with any neutral vector meson in the final state, as well). The region denoted $R_{III}$ is accessible to the muon and the $\rho$ meson of the three- and four-body decays, while $R_{IV-III}$ is allowed only for the corresponding radiative decay. The evaluation of the integrals in Eq. (\ref{bremss1}) is done by considering appropriately the region of phase space corresponding to three- and four-body regions of the Dalitz plot.

\subsubsection{Model-dependent corrections}
\begin{figure}\centering
\includegraphics[scale=0.55]{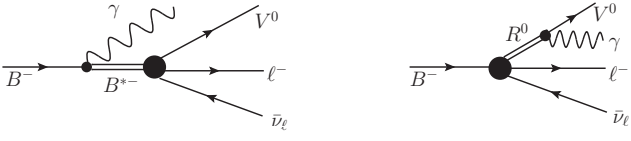}
\caption{\footnotesize Model-dependent real photon corrections; $R^0$ denotes a pseudoscalar resonance.}
\label{moddepfig}
\end{figure}

In addition to the amplitude ${\cal M}^{\rm Low}$ for real photon emission obtained by attaching the photon in all possible ways, one can have also model-dependent terms which are of $O(k)$ and higher and are contained in ${\cal M}^{MD}$. For relatively low photon energies,  it has been pointed out  \cite{Becirevic} that the diagrams shown in Figure \ref{moddepfig} must be taken into account in $B\to D\ell \nu_{\ell}\gamma$ decays since on-shell $D^*(\to D\gamma)$ intermediate states can have an important contribution as non-radiative events. For the decay channels considered in this paper $B^{\pm}\to (\rho^0, \omega, D^{*0})\ell^{\pm}\nu_{\ell}$, there are not nearby narrow intermediate states decaying into $V^0\gamma$ states that can produce a similar important effect. 

For hard photons, the structure of electromagnetic vertices in terms of elementary components becomes important and an approach based on QCD degrees of freedom like the Soft Collinear Effective Theory \cite{Cirigliano2006} looks more appropriate. Our evaluation of the model-dependent contributions to ${\cal M}^{MD}$ for moderately energetic photons follows closely Ref. \cite{Becirevic}. 

The general form of the MD decay amplitude becomes
\begin{eqnarray}\label{md1}
{\cal M}^{MD} = e \frac{G_F}{\sqrt{2}} V_{qb} c_V \epsilon^{\mu}~(V_{\mu\nu}^{MD}-A_{\mu\nu}^{MD}) L^{\nu} \ .
\end{eqnarray}
From Figure \ref{moddepfig} we get \cite{Becirevic}
\begin{eqnarray}\label{VAMD}
V_{\mu\nu}^{MD} &-& A_{\mu\nu}^{MD} =\nonumber \\
&& \frac{ \left\langle V^0 |(V_{\nu}-A_{\nu})|B^{*-}(P_1)\right\rangle \left\langle B^{*-}(P_1)|J_{\mu}|B^- \right\rangle}{P_1^2 -m_{B^{*}}^2+i\epsilon}  \nonumber \\
&&+\frac{ \left\langle V^0 |J_{\mu}|R^0(P_2)\right\rangle \left\langle R^0(P_2)|(V_{\nu}-A_{\nu})|B^- \right\rangle}{P_2^2 -m_{R}^2+i\epsilon} ~.
\end{eqnarray}
where $P_1=P-k$ and $P_2=P_V+k$ denote the four momenta of the corresponding resonances, and $m_{B^{*}}(m_R)$ the mass of $B^{*-}(R^0)$ (if the $R$ resonance is produced on-shell, we have to use $\epsilon\to m_R\Gamma_R$).  In the above expression $J_{\mu}$ denotes the electromagnetic current and $(V_{\nu}-A_{\nu})$ denotes the usual V-A weak currents \cite{Becirevic}. 

We define the transition matrix elements of the electromagnetic current as 
\begin{eqnarray}\label{emc}
\left\langle B^{*-}(P_1)|J_{\mu}|B^- \right\rangle &=& i g_{B^- B^{*-} \gamma} \epsilon_{\mu \alpha \beta \eta} \eta^{\alpha} P_1^{\beta} k^{\eta} \\ \nonumber
\left\langle V^0 |J_{\mu}|R^0(P_2)\right\rangle &=& i g_{R^0 V^0 \gamma} \epsilon_{\mu \alpha \beta \eta} \varphi^{\alpha} P_2^{\beta} k^{\eta} ~,
\end{eqnarray}
with $g_{PV\gamma}$ denoting the $VP\gamma$ coupling, and $\eta$ the polarization four vector of $B^{*}$. As expected, the MD amplitudes in Eq. (\ref{md1}) are of $O(k)$; as it was noticed in Ref. \cite{Becirevic}, when the intermediate resonances become on-shell the corresponding decay amplitude in (\ref{VAMD}) become of $O(k^0)$. In our case, we do not expect MD contributions with on-shell resonances, although it is interesting to evaluate the effect given that in the soft-photon limit some contributions behave as $(m_V^2-m_P^2)^{-1}$.
   
The $B\to R$ weak matrix element is defined as usual
\begin{eqnarray}\label{bpi}
\left\langle R(P_V)|(V_{\nu} \right. &-& \left. A_{\nu})|B^-(P) \right\rangle = \nonumber \\
&& \left[(P+P_V)_{\nu}-\frac{M^2-m_R^2}{q^2} q_{\nu} \right]F_1(q^2) \nonumber \\ 
&& + \frac{M^2-m_R^2}{q^2} q_{\nu} F_0(q^2) ~,
\end{eqnarray}
with $F_{1,0}$ the vector and scalar form factors, respectively.  We have neglected the $k$ dependence in evaluating the matrix element since this would induce ${\cal O}(k^2)$ terms in the MD amplitude, which are not  considered in our approximation.

Since the main decay of the $B^*$ meson is electromagnetic ($B^*\to B\gamma$), there are not available calculations of its weak $B^*\to V$ matrix element. Thus, we have to rely on the Heavy Quark Symmetry \cite{Isgur} to estimate the  $B^{*-}\to V^0$ weak transition. This symmetry allows to relate this matrix element  to the $B^{-}\to V^0$ matrix element, in the limit of an infinitely heavy $b$ quark. 
Using the algebra of heavy quarks operators in the HQET \cite{Neubert}, one can derive:
\begin{eqnarray}\label{bstar}
\left\langle V^0 |\bar{q}\gamma^{\mu}(1\right.&-&\left.\gamma_5)Q|B^{*-}(v,\eta)\right\rangle = \nonumber \\ 
&& \ \ \ \left\langle V^0 |\bar{q}\gamma_{\beta}(1-\gamma_5)Q|B^-\right\rangle \nonumber \\
&& \ \ \ \times i[\epsilon^{\mu\nu\alpha\beta}v_{\nu}\eta_{\alpha}-i(\eta^{\mu}v^{\beta}-v^{\mu}\eta^{\beta})] \ ,
\end{eqnarray} 
where $v=P_B/m_B$ is the four-velocity of the heavy meson and $\eta$ the polarization four-vector of the $V^0$ vector meson. Since we are interested in getting an estimate of this model-dependent contribution  to the real-photon radiative correction of $O(\alpha)$, this approximation should be good enough.

In the following subsection, we define the different contributions to the radiative corrections. As noticed before, we call the model-dependent (MD) part of the radiative amplitude to the sum of the last two terms in Eq. (\ref{bremss}) and the amplitude defined in Eqs. (\ref{md1})-(\ref{VAMD}).

\subsection{Order $\alpha$ radiative corrections}

By including the effects of LD virtual and real photon corrections that were obtained in Eq. (\ref{wvirtual}) and section 2,   we can write  the following expression for the photon-inclusive Dalitz plot distribution 
\begin{eqnarray}\label{diffrate}
\frac{{\rm d}^{2}{\Gamma}}{{\rm d}E{\rm d}E_{\rho}} &=& \frac{{\rm d}^{2}{\Gamma}^0}{{\rm d}E{\rm d}E_{\rho}} \left[\frac{}{}1 + \tilde{\delta}^1_{LD}(E,E_{\rho},\Lambda) \right] \nonumber \\ && + \frac{{\rm d}^{2}\Gamma^{MI}_{IV-III}(E,E_{\rho})}{{\rm d}E{\rm d}E_{\rho}} \nonumber \\ 
&& \frac{{\rm d}^{2}\Gamma^{MI-MD}(E,E_{\rho})}{{\rm d}E{\rm d}E_{\rho}} + \frac{{\rm d}^{2}\Gamma^{MD}(E,E_{\rho})}{{\rm d}E{\rm d}E_{\rho}}\ .
\end{eqnarray} 
The first term in the above expression, $\tilde{\delta}^1_{LD}(E,E_{\rho},\Lambda^{2}) $, contains the effects of virtual photons and bremsstrahlung contributions of (Eq. \ref{bremss1}) that should be integrated in the phase-space region $R_{III}$ of $(E,E_V)$  compatible with the kinematics of non-radiative (three-body) decays. This piece of radiative corrections is infrared finite, model-independent \cite{Sirlin}, but depend upon the UV cutoff scale $\Lambda$ that determines the range of validity of the LD approximation, as it was mentioned at the beginning of this section. 
 
\begin{figure}\centering
\includegraphics[scale=0.3]{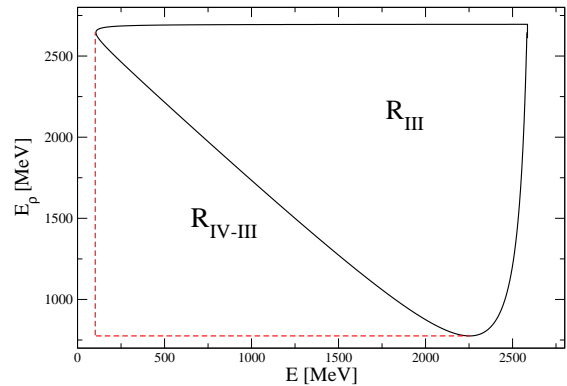}
\caption{\footnotesize Phase-space region accessible to $B^- \to \rho^0 \mu^-\nu_{\ell}(\gamma)$ decays. The region $R_{III}$ is allowed to three- and four-body (radiative) decays, while $R_{IV-III}$ is accesible only in the radiative mode.}
\label{dalitz}
\end{figure}

 The second term in Eq. (\ref{diffrate}) contains the contribution of the MI terms of the real-photon corrections of Eq. (\ref{bremss1}) evaluated in the $R_{IV-III}$ region of phase space (see Figure \ref{dalitz}) \cite{Ginsberg,Alain,cirigliano2008}. The last two terms in Eq. (\ref{diffrate}) arise from the interference of model-independent and model-dependent ($MI-MD$), and the square of the model-dependent ($MD$) contributions, respectively, integrated over the full kinematical range. Let us note that we have included Eq. (\ref{md1}) and the last two terms of Eq. (\ref{bremss}) as the components of the MD amplitude. 

On the other hand, the short-distance corrections (photon momenta above $\Lambda$) can be calculated using the electroweak  theory in terms of elementary field components. At $O(\alpha)$, the dominant logarithmic term of this correction to the decay rate is given by Eq. (\ref{sdcorrection}) \cite{sirlin1978}, where the mass of the $Z$ boson serves as a natural UV cutoff in the EW theory. As  is well known, the cutoff-dependence does not cancel completely in the sum of $\tilde{\delta}^1_{LD}$ and $\delta^1_{SD}$ corrections. Actually, the $\Lambda$-dependent term remaining in the sum of LD and SD corrections, $\Delta\delta^1(\Lambda)=\alpha \ln(m_Z/\Lambda)/(2\pi)$ changes from $0.63\%$ to $0.33\%$ when $\Lambda$ varies from 0.4 GeV up to the $B$ meson mass scale. Note that the $\Lambda$-dependence can be eliminated, as done for instance, in the calculation  of radiative corrections to $K_{\ell 3}$ \cite{Cirigliano} and semileptonic tau decays \cite{Cirigliano1} within the framework of chiral perturbation theory; there, a local counterterm can be chosen and added to the LD contributions in order to achieve a complete matching between LD and SD corrections. Within the approach used in the present paper, we will consider the effect of this cutoff dependence upon the observable decay rate as an uncertainty of the calculation.

\section{Results}

In this section we present the results of radiative corrections to the rates of $B^- \to V^0 \ell^- \bar{\nu}_{\ell}$ decays, where $V^0$ is a neutral vector meson and $\ell$ a charged lepton. The integrals involved in virtual and real photon contributions were evaluated numerically. 

\subsection{Corrections to $B^- \to (\rho^0, \omega) \ell^-\bar{\nu}_{\ell}$ decays}

 In this case we  use the form factors of the $B\to V$ transitions as provided by lattice \cite{FF} and light-cone QCD sum rules \cite{lcsr} calculations, in the cases of $V=\rho^0,\omega$ mesons. In the full range of $q^2$ values, the LD radiative corrections turned out to be rather insensitive to these two different models for the form factors; we will assign an uncertainty due to the numerical evaluation and to account for the different yields obtained by using the above parametrizations for the form factors. For the input required in model-dependent terms, we use the form factors $F_{1,0}^{B\to \pi}(q^2)$ given in \cite{BCL}, and the following values of coupling constants for $VP\gamma$ vertices: $g_{B^- B^{*-} \gamma}=(1.0\pm 0.1)\times 10^{-3}$ MeV$^{-1}$ \cite{Becirevic2} and $g_{\pi \rho \gamma}=(7.3\pm0.8)\times 10^{-4}$ MeV$^{-1}$ \cite{Alain2}.

Table \ref{tablerho} shows the results for the short-distance (SD), long-distance (LD) and the total radiative corrections $\delta_{T}^1(\ell)$, at three different values of the scale $\Lambda$, for the $B^- \to \rho^0 \ell^- \bar{\nu}_{\ell}$ decay rates. From this table we quote the following radiative corrections to the total decay: 
\begin{eqnarray}\label{radcorrho0}
\delta_{T}^1(\ell)= \left\{ \begin{array}{ll} (1.62\pm 0.10\pm 0.04) \%\ ,  & {\rm for}\ \ell=\mu  \cr
(1.63 \pm 0.11\pm 0.04)\% \ ,& {\rm for}\ \ell=e \end{array} \right. \ .
\end{eqnarray}
We have taken as central value for $\delta^1_{T}(\ell)$ the result corresponding to the  cutoff  scale $\Lambda=m_{\rho}$. The first (larger) error is obtained by varying $\Lambda$ in the range  shown in Table \ref{tablerho}, while the second one is associated to the use of two different parametrizations of the form factors \cite{FF,lcsr} and the numerical evaluation.  
In  the absence of a rigorous prescription to compute radiative corrections in the intermediate energy regime that gives a perfect matching between scalar QED  and the Standard Electroweak theory, we have to deal with this rather small uncertainty.

\begin{table}[t!]\caption{\footnotesize Contribution of the short-distance (SD) and long-distance (LD) radiative corrections to $B^-\rightarrow \rho^0 \ell^- \bar{\nu}_{\ell}$  ($\ell=\mu$ or $e$) at three different matching scales. The LD corrections include contributions from three- and four-body regions, which partially cancel each other.}\label{tablerho}\scriptsize
\begin{tabular}{|c|c||c|c||c|c|}
\hline 
$\Lambda$ & $\delta^1_{SD}$ & $\delta^1_{LD}(\mu)$ & $\delta^1_{T}(\mu)$ & $\delta^1_{LD}(e)$ & $\delta^1_{T}(e)$\\ 
\hline 
$m_{\rho}/2$ \ & \ 0.0254 & \ -0.0085 \ & \ 0.0168\ & \ -0.0083\ & 0.0171\ \\ 
\hline 
$m_{\rho}$ \ & \ 0.0221 & \ -0.0060 \ & \ 0.0162\ & \ -0.0058 \ & 0.0163\ \\ 
\hline 
$2$ GeV & \ 0.0177 \ & \ -0.0025 \ & \ 0.0152 \ & \ -0.0025 \ & 0.0152\ \\ 
\hline 
\end{tabular}
\end{table}

A few comments are in order: 
\begin{enumerate}
\item As it can be observed from Table \ref{tablerho}, the LD corrections turn out to be very small, at the few per mille level. This is the case because we are including the effects of radiative corrections coming from radiative events falling in the four-body region, which almost cancel the LD contributions coming from the three body region. Actually, events from the R$_{IV-III}$ region, see Fig. \ref{dalitz}, contribute $\delta^1_{LD}({\rm R}_{IV-III}) = +0.0096$ $(+0.0352)$ to the LD corrections in Table \ref{tablerho} of the rates into a muon (electron) channels.  This trend is in agreement with similar conclusions gotten in Ref. \cite{cirigliano2008} in the case of  $K^-_{\ell 3}$ decay. 

\item Inclusion of model-dependent contributions gives a correction $\sim -0.02\%$, an order of magnitude smaller than the model independent ones, for both leptons. The smallness of such a contribution is due to a partial cancellation between interference ($MI-MD$) and model-dependent ($MD$) corrections. On the one hand, the correction associated to $MI-MD$, $\sim -0.07\%$, is dominated by the interference of the last two terms of the Low's amplitude with the first line of Eq. (\ref{bremss}). On the other hand, the MD correction, $\sim +0.05\%$, is dominated by the MD amplitude in Eqs. (\ref{md1})-(\ref{VAMD}). Each of these corrections are found to contribute at the per mille level, as was noticed in $K_{\ell 3}$ decays \cite{Andre,ruben1}, but the relative sign between them makes the final result even smaller.        

\item Recently, bremsstrahlung simulations for specific channels in $K_{\ell 3}$ decays were implemented in PHOTOS  \cite{PHOTOS,PH2012}. This novelty incorporates terms of $O(k^0)$ in the decay amplitude which were found to be negligible compared to the contribution of the universal emission kernel  \cite{PHOTOS,PH2012}. The approximation considered in Refs.  \cite{PHOTOS,PH2012} would correspond to neglecting the last term in Eq. (\ref{bremss}) in the amplitude of $B^+\to V^0$ transitions considered in this paper. When we compare the results of LD corrections obtained using our full calculation with the one in PHOTOS' approximation, we get a difference of $+0.05\%$ for both leptonic channels.

\item Given the equality of the matrix elements in Eq. (\ref{isospin}) and the near equality of $\rho^0$ and $\omega$ meson masses, we expect the radiative corrections to $B^- \to \omega(782)\ell^-\bar{\nu}_{\ell}$ to be the same as in Eq. (\ref{radcorrho0}). An explicit numerical evaluation based on the previous formulae, confirms this expectation. Thus, we conclude that the dominant source of isospin breaking to the second of the equalities in Eq. (\ref{isospin}) should be given by the effect of $\rho-\omega$ mixing \cite{rho-omega}, and not by radiative corrections. Note that the ratio of  $(B^- \to \rho^0)/(B\to \omega)$ semileptonic rates is rather independent of the cutoff scale $\Lambda$ and form factor models of the $B^- \to V^0$ transition.
\end{enumerate}

For completeness, we provide also the evaluations of the LD radiative corrections to the Dalitz plot distribution. These corrections can be useful for experimental studies aiming a precise extraction of the weak form factors \cite{Ginsberg,Garcia,ruben1} which focus in the three-body region of the phase space. For this purpose, we re-write Eq. (\ref{diffrate}) as follows: 
\begin{equation}\label{dpdist}
\frac{{\rm d}^2\Gamma}{{\rm d}E{\rm d}E_{\rho}}=\frac{{\rm d}^2\Gamma^0}{{\rm d}E{\rm d}E_{\rho}}\left[ 1+\Delta^1_{LD}(E,E_{\rho}) \right]\ ,
\end{equation}
with an obvious definition of the LD radiative correction function $\Delta^1_{LD}(E,E_{\rho})$ to the Dalitz plot. 
In Table \ref{DPrho} we show the values of the $\Delta^1_{LD}(E,E_{\rho})$ function for different energies of final states particles in the three-body region of the phase space. The  large and negative values of radiative corrections for more relativistic charged leptons tend to dominate the contribution of the three-body phase space region of radiative corrections. Despite the similar values of the total radiative corrections for both leptons, it is clear from Table \ref{DPrho} that the corrections can vary up to an order of magnitude from one leptonic channel to the other at the same point of the Dalitz plot.
\begin{table*}
 \caption{\footnotesize LD radiative correction function $\Delta_{LD}^1(E,E_{\rho})$ to the $B^- \to \rho^0 e^- \bar{\nu}_{e}$ (upper half) and  $B^- \to \rho^0 \mu^- \bar{\nu}_{\mu}$ (lower half) Dalitz plot distributions  ($\Lambda = m_{\rho}$). The energies $E$ and $E_{\rho}$ are given in MeV units.}\label{DPrho}
\begin{tabular}{|c||c|c|c|c|c|c|c|c|c|c|}\hline 
$E_{\rho}~\setminus~E $& 200 & 450 & 700 & 950 & 1200 & 1450 & 1700 & 1950 & 2200 & 2450\\ \hline \hline
2650 & 0.1001 & 0.0501 & 0.0251 & 0.0071 & -0.0084 & -0.0230 & -0.0390 & -0.0560 & -0.0800 & -0.1200 \\ \hline
2450 &   & 0.0422 & 0.0232 & 0.0072 & -0.0073 & -0.0219 & -0.0369 & -0.0549 & -0.0790 & -0.1300 \\ \hline
2250 &   &  & 0.0183 & 0.0040 & -0.0094 & -0.0228 & -0.0378 & -0.0559 & -0.0799 & -0.1300 \\ \hline
2050 &   &  & 0.0143 & 0.0010 & -0.0117 & -0.0247 & -0.0397 & -0.0568 & -0.0809 & -0.1300 \\ \hline
1850 &   &  &  & -0.0019 & -0.0136 & -0.0267 & -0.0407 & -0.0578 & -0.0819 & -0.1299 \\ \hline
1650 &   &  &  &  & -0.0157 & -0.0287 & -0.0427 & -0.0588 & -0.0839 & -0.1399 \\ \hline
1450 &   &  &  &  &  & -0.0308 & -0.0438 & -0.0609 & -0.0849 & -0.1399 \\ \hline
1250 &   &  &  &  &  &  & -0.0459 & -0.0630 & -0.0880 & -0.1499 \\ \hline
1050 &   &  &  &  &  &  &  & -0.0652 & -0.0901 & -0.1699 \\ \hline
850 &   &  &  &  &  &  &  &  & -0.0964 &  \\ \hline \hline 
2650 & -0.0200 & 0.0035 & 0.0007 & -0.0032 & -0.0073 & -0.0120 & -0.0170 & -0.0220 & -0.0300 & -0.0450 \\ \hline
2450 &   & 0.0078 & 0.0044 & 0.0002 & -0.0041 & -0.0089 & -0.0139 & -0.0199 & -0.0290 & -0.0460 \\ \hline
2250 &   &  & 0.0041 & 0.0000 & -0.0043 & -0.0090 & -0.0138 & -0.0209 & -0.0289 & -0.0460 \\ \hline
2050 &   &  & 0.0035 & -0.0005 & -0.0047 & -0.0093 & -0.0147 & -0.0208 & -0.0299 & -0.0470 \\ \hline
1850 &   &  &  & -0.0010 & -0.0051 & -0.0097 & -0.0147 & -0.0208 & -0.0299 & -0.0479 \\ \hline
1650 &   &  &  &  & -0.0056 & -0.0097 & -0.0147 & -0.0218 & -0.0299 & -0.0489 \\ \hline
1450 &   &  &  &  &  & -0.0108 & -0.0158 & -0.0219 & -0.0309 & -0.0509 \\ \hline
1250 &   &  &  &  &  &  & -0.0159 & -0.0230 & -0.0320 & -0.0539 \\ \hline
1050 &   &  &  &  &  &  &  & -0.0232 & -0.0331 & -0.0619 \\ \hline
850 &   &  &  &  &  &  &  &  & -0.0344 &  \\ \hline
\end{tabular}
\end{table*}

\subsection{Corrections to the $B^- \to D^{*0} \ell^- \bar{\nu}$ rate}

We can apply the same formalism to compute the LD QED radiative corrections to the decay $B^{-}\rightarrow D^{*0} \ell^- \bar{\nu}_{\ell}$, by replacing the corresponding masses and hadronic form factors.  For the numerical evaluation of virtual and model-independent real-photon corrections we use the form factor parametrization given in \cite{caprini} and the results of fits obtained in refs. \cite{milc,hfag}. The relation between those parametrizations and the form factors in Eq. (\ref{hme}) are given in \cite{fajfer}. In addition, for model-dependent real-photon contributions, we use the value $g_{D^0 D^{*0} \gamma}=(1.0\pm 0.9)\times 10^{-3}$ MeV$^{-1}$ for the coupling constant extracted from $D^{*0}$ and $D^{*\pm}$ decays \cite{pdg2014}. The form factors $F_{1,0}^{B \to D}$ are taken from Ref. \cite{BtoDff}. Since photons are not very hard in this case ($E_{\gamma}^{\rm max}\approx 1.28$ GeV), our approximations for the real-photon amplitude should be more reliable.

Table \ref{tableD} summarizes our results for the radiative corrections to $B^{-}\rightarrow D^{*0} \ell^- \bar{\nu}_{\ell}$, from which we can extract
\begin{eqnarray}
\delta^1_{T}(\ell)= \left\{ \begin{array}{ll}(1.53 \pm 0.06 \pm 0.04)\%, & {\rm for}\  \ell=\mu \cr
(1.53 \pm 0.08 \pm 0.04)\%, & {\rm for}\  \ell=e \end{array} \right. \ .
\end{eqnarray} 
As in the previous case, the first uncertainty corresponds to variations of the  (un)matching scale $\Lambda$ and the second one to the use of different form factor models and the numerical evaluation. Similar to the previous case, both corrections turn out to be equal for both leptonic channels. In this case, we chose (arbitrarily) the central value for the matching scale $\Lambda = m_{D^{*0}}$, in analogy to ref. \cite{bernlochner} in the case of (pseudo)scalar charmed meson, but we have included the range of $\Lambda$ within $m_{D^{*0}}/2$ and $2m_{D^{*0}}$.

\begin{table}[t!]\caption{\footnotesize Same description as in Table \ref{tablerho} for $B^- \to D^{*0}\ell^-\nu_{\ell}$ decay rates}\label{tableD}\scriptsize
\begin{tabular}{|c|c||c|c||c|c|}
\hline 
$\Lambda$ & $\delta^1_{SD}$ & $\delta^1_{LD}(\mu)$ & $\delta^1_{T}(\mu)$ & $\delta^1_{LD}(e)$ & $\delta^1_{T}(e)$\\ 
\hline 
$m_{D^{*0}}/2$ \ & \ 0.0209 & \ -0.0051 \ & \ 0.0159\ & \ -0.0048\ & \ 0.0161\ \\ 
\hline 
$m_{D^{*0}}$ \ & \ 0.0177 & \ -0.0024 \ & \ 0.0153\ & \ -0.0024\ & \ 0.0153\ \\ 
\hline 
$2m_{D^{*0}}$ & \ 0.0145 \ & \ 0.0003 \ & \ 0.0148 \ & \ 0.0001 \ & \ 0.0146\ \\ 
\hline 
\end{tabular}
\end{table} 
 
As in the case of charmless $B^- \to V^0$ transitions, a partial cancellation occurs between LD corrections evaluated  in the three- and four-body regions of phase space. In this case, the LD effects from the R$_{IV-III}$ region of the Dalitz plot contribute $+0.0051$ and $+0.0223$ to the radiative corrections into muon and electron channels in Table \ref{tableD}, respectively. It is worth to mention that such cancellation of LD corrections has been noticed previously  in radiative corrections to $K^{-}\rightarrow \pi^0 \ell^- \bar{\nu}_{\ell}$ decays \cite{cirigliano2008}. The interference term $MI-MD$ gives a correction $\sim -0.10\%$, while the $MD$ term $\sim +0.06\%$, with the final result $\sim -0.04\%$, for the two leptonic channels; the origin of such values is explained in the previous subsection. As in the case of $B^+\to (\rho^0,\omega)$ transitions, the difference between our full calculation of the LD corrections and the one obtained in the PHOTOS' approximation \cite{PHOTOS,PH2012} is $+0.05\%$ in the integrated rate.

In Table \ref{DPdstar} we show the contribution of LD radiative correction function $\Delta^1_{LD}(E,E_{D^*})$, defined in Eq. (\ref{dpdist}),  to the Dalitz plot distributions of $B^- \to D^{*0}\ell^- \bar{\nu}_{\ell}$ decays.
Similarly, the predominance of radiative corrections of negative sign in most of the three-body region of phase space turns out to give radiative correction of negative sign for the LD corrections contributions to the decay rate coming from the R$_{III}$ region of the Dalitz plot. 
\begin{table*}
\caption{\footnotesize Same description as in Table \ref{DPrho} for the cases of $B^- \to D^{*0} e^-\bar{\nu}_e$ (upper half) and $B^- \to D^{*0} \mu^-\bar{\nu}_{\mu}$ (lower half) decays.}\label{DPdstar}
\begin{tabular}{|c||c|c|c|c|c|c|c|c|c|c|}\hline 
$E_{D^{*0}}~\setminus~E $& 200 & 400 & 600 & 800 & 1000 & 1200 & 1400 & 1600 & 1800 & 2000\\ \hline \hline
3000 &  0.0931 & 0.0501 & 0.0271 & 0.0111 & -0.0034 & -0.0170 & -0.0300 & -0.0450 & -0.0620 & -0.0870 \\ \hline
2900 &  0.0772 & 0.0484 & 0.0294 & 0.0144 & 0.0003 & -0.0127 & -0.0268 & -0.0418 & -0.0599 & -0.0859 \\ \hline
2800 &   & 0.0435 & 0.0266 & 0.0126 & -0.0005 & -0.0135 & -0.0266 & -0.0417 & -0.0598 & -0.0859 \\ \hline
2700 &   & 0.0384 & 0.0236 & 0.0107 & -0.0016 & -0.0144 & -0.0275 & -0.0416 & -0.0607 & -0.0869 \\ \hline
2600 &   &  & 0.0216 & 0.0093 & -0.0027 & -0.0143 & -0.0274 & -0.0426 & -0.0607 & -0.0889 \\ \hline
2500 &   &  &  & 0.0079 & -0.0038 & -0.0153 & -0.0284 & -0.0436 & -0.0617 & -0.0919 \\ \hline
2400 &   &  &  & 0.0065 & -0.0048 & -0.0163 & -0.0294 & -0.0446 & -0.0638 & -0.0989 \\ \hline
2300 &   &  &  &  & -0.0058 & -0.0174 & -0.0305 & -0.0456 & -0.0658 & -0.1200 \\ \hline
2200 &   &  &  &  &  & -0.0185 & -0.0316 & -0.0477 & -0.0709 &  \\ \hline
2100 &   &  &  &  &  &  & -0.0337 & -0.0518 & -0.0870 &  \\ \hline \hline
3000 &  0.0032 & 0.0042 & 0.0014 & -0.0019 & -0.0052 & -0.0087 & -0.0120 & -0.0170 & -0.0220 & -0.0300 \\ \hline
2900 &  0.0140 & 0.0113 & 0.0080 & 0.0042 & 0.0004 & -0.0036 & -0.0081 & -0.0128 & -0.0189 & -0.0279 \\ \hline
2800 &   & 0.0124 & 0.0085 & 0.0047 & 0.0009 & -0.0032 & -0.0076 & -0.0127 & -0.0188 & -0.0289 \\ \hline
2700 &   & 0.0123 & 0.0084 & 0.0048 & 0.0009 & -0.0031 & -0.0076 & -0.0126 & -0.0187 & -0.0289 \\ \hline
2600 &   &  & 0.0082 & 0.0046 & 0.0009 & -0.0032 & -0.0077 & -0.0126 & -0.0197 & -0.0289 \\ \hline
2500 &   &  &  & 0.0045 & 0.0007 & -0.0032 & -0.0077 & -0.0126 & -0.0197 & -0.0299 \\ \hline
2400 &   &  &  & 0.0042 & 0.0006 & -0.0034 & -0.0079 & -0.0136 & -0.0198 & -0.0319 \\ \hline
2300 &   &  &  &  & 0.0003 & -0.0037 & -0.0082 & -0.0136 & -0.0208 & -0.0380 \\ \hline
2200 &   &  &  &  &  & -0.0040 & -0.0086 & -0.0137 & -0.0229 &  \\ \hline
2100 &   &  &  &  &  &  & -0.0093 & -0.0158 & -0.0280 &  \\ \hline
\end{tabular}
\end{table*}

\section{Conclusions}
We have calculated the long-distance (LD) corrections to the charmless and charmfull semileptonic $B^- \to V^0 \ell^- \bar{\nu}_{\ell}$ decays  involving the ground state vector mesons $V^0$ and  $\ell=\mu$ and $e$ flavored leptons.  We have used a formalism where the LD corrections are approximated by scalar QED, and assumed valid up to a  scale $\Lambda$, which serves as a separation cutoff of LD and short-distance (SD) corrections. Under these assumptions, the cutoff dependent corrections of $O(\alpha)$ depend mildly upon the scale $\Lambda$. We have associated an uncertainty to radiative corrections in the decay rate due to this cutoff dependence by allowing it to vary between $m_V/2$ and approximately $2m_V$. Also, in agreement with a similar analysis for $K_{\ell 3}$ decays \cite{Andre}, the effects of using  different models for the weak form factors are estimated to play a subleading role compared to the dependence upon the separation scale $\Lambda$. 

 The use of scalar QED in the hadronic vertices  may be justified for relatively low values of the cutoff $\Lambda$ in the photon momenta. In the case of real-photon corrections we have included, in addition to the Low's soft-photon amplitude, some model-dependent contributions that originate in the exchange of meson resonances. This allows   to cover harder real photons that are suitable to describe photon-inclusive rates. These model-dependent contributions, are found to be ten times smaller than the one obtained from the soft-photon approximation. 

We have found that there exist large cancellations between the contributions of LD corrections to the decay rates coming from the three- and four-body regions of the Dalitz plot. These cancellations occurs for the integrated rates of photon-inclusive $B^- \to V^0$ transitions. This happens for both, charmless and charmfull transitions, and for both flavors of leptons. This is in agreement with previous calculations of radiative corrections to $K^{\pm}_{3\ell}$ decay rates \cite{cirigliano2008}. 

Our calculation of virtual corrections assumes a single separation scale $\Lambda$ of LD and SD corrections long-distance, which avoids the consideration of structure-dependent effects of photon-hadron interactions in the region populated by resonances. Althought the calculation of intermediate distance effects deserves further studies (beyond the scope of this paper), currently we miss a precise prescription that matches the calculations based on the fundamental and the effective low-energy theories used to compute, respectively, the SD and LD radiative corrections. At present, any calculation can be based at most in a model-dependent approach \cite{Decker}.

Our results can be useful for future/improved measurements of the different charged decay channels of $B\to V \ell^- \bar{\nu}_{\ell}$ branching ratios in order to test consistency of data with isospin symmetry. To complete this test, the calculation of radiative corrections to $\bar{B}^0 \to V^+\ell^-\bar{\nu}_{\ell}$ (and remaining isospin breaking corrections in form factors) will be necessary, which we plan as a future project since the electromagnetic vertex of the charged vector meson requires a careful treatment. Actually, the determination of the $V_{ub}$ and $V_{cb}$ matrix elements at the few percent level, requires the consideration of the full radiative corrections of $O(\alpha)$. 

\

{{\bf Acknowledgements}: 
 The authors would like to thank Conacyt (M\'exico) for partial financial support under projects  296 (Fronteras de la Ciencia), 236394 and 263916. This work has been supported in part by EPLANET project and by the Spanish Government and ERDF funds from the EU Commission
[Grants No. FPA2011-23778, FPA2014-53631-C2-1-P, SEV-2014-0398] and by Generalitat Valenciana under Grant No. PROMETEOII\break /2013/007. The authors are indebted to Ruben Flores-Mendieta and Pablo Roig for useful discussions and comments.

\

\appendix

\
\section{Relevant scalar integrals}

The scalar two- and three-point functions in the cutoff approximation are given by
\begin{equation}\label{stpf}
{\rm B}_0[m^{2},0,m^{2}] = 2 - {\ln}\left(\frac{m^2}{\Lambda^{2}} \right)
\end{equation}

\begin{eqnarray}\label{Blm}
{\rm B}^{lM}_{0}[u,m^2,M^{2}] &=&  \nonumber \\ && \!\!\!\!\!\!\!\!\!\!\!\!\!\!\!\!\!\!\!\!\!\!\!\!\!\!\!\!\!\!\!\!\!\!\!\!\!\!\!\! - \int_{0}^{1} {\rm d x}{\ln}\left(\frac{-x(1-x)u + x M^{2}+(1-x)m^2}{\Lambda^{2}}  \right) \ ,
\end{eqnarray}
where we have used $\Lambda$ as the UV-cutoff in LD corrections. The derivative of Eq. (\ref{stpf}) with respect to $p^2$ gives 
\begin{equation}\label{derivative}
{\rm B}'_0[m^2,\lambda^{2},m^{2}] = - \frac{1}{2m^2}\left(2 + {\ln}\left(\frac{\lambda^{2}}{m^2} \right) \right) \ ,
\end{equation}
where $\lambda$ represents a ficticious photon mass, introduced to regulate the IR-divergence. Similar expressions to Eqs. (\ref{stpf}),(\ref{derivative}) can be obtained for the scalar integrals associated to the loops involving the charged meson by proper replacements of  momenta and masses.

The function C$_0$ can be split into two terms
\begin{eqnarray}
{\rm C}_{0}[m^2,M^2,u,m^2,\lambda^{2},M^{2}] &=& -\frac{1}{4 M E \beta } {\ln}\left(\frac{1+\beta}{1-\beta} \right) \nonumber \\ 
&&\!\!\!\!\!\!\!\! \times {\rm ln}\left(\frac{u}{\lambda^{2}} \right)
 - {\rm F}_2(P,p)  \ ,
\end{eqnarray}
where $\beta= |\mathbf{p}|/E$ is the charged lepton velocity in the rest frame of the decaying particle. The first term is IR-divergent while the second is IR-finite and given by
\begin{equation}
{\rm F}_2(P,p)  = \frac{1}{2} \int_{0}^{1} {\rm d y} \ \frac{1}{\chi(y)}{\ln}\left(\frac{\chi(y)}{u} \right) \ ,
\end{equation}
with $\chi(y) = uy^2 + 2(p\cdot P - p^{2})y + p^2 $.

The three-point functions ${\rm C}_{1,2}$ can be reduced to scalar two point integrals as
\begin{eqnarray}
{\rm C}_{1}[m^2,M^2,u,m^2,0,M^{2}] &=& -  \frac{1}{2 M^2 \beta^{2}}  \left[\frac{}{} p^2 \left( {\rm B}_0[m^{2},0,m^{2}] \right. \right. \nonumber \\ && 
\!\!\!\!\!\!\!\!\!\!\!\!\!\!\!\!\!\!\!\!\!\!\!\! \!\!\!\!\!\!\!\!\!\!\!\!\!\! \!\!\!\!  \left.  \left. -{\rm B}^{lM}_0[u,m^2,M^{2}] \right)  
\!\!  - P\cdot p \left({\rm B}^{M}_0[M^{2},0,M^{2}] \right. \right. \nonumber \\ && 
\left. \left. !\!\!\!\!\!\!\!\!\!-{\rm B}^{lM}_0[u,m^2,M^{2}] \right) \frac{}{}\right]
\end{eqnarray}
\begin{eqnarray}
{\rm C}_{2}[m^2,M^2,u,m^2,0,M^{2}] &=& -  \frac{1}{\textstyle 2 M^2 \beta^{2}} \nonumber \\
&& \!\!\!\!\!\!\!\!\!\!\!\!\!\!\!\!\!\!\!\!\!\!\!\!\!\!\!\!\!\!\!\!\!\!\!\!\!\!\!\!\!\!\!\!\!\!\!\!\!\!\!\!\!\!\! \times \left[\frac{}{}\!\!\! - P\cdot p \left({\rm B}_0[m^{2},0,m^{2}]  -{\rm B}^{lM}_0[u,m^2,M^{2}] \right) \right. \nonumber \\
&& \left. \!\!\!\!\!\!\!\!\!\!\!\!\!\!\!\!\!\!\!\!\!\!\!\!\!\!\!\!\!\!\!\!\!\!\!\!\!\!\!\!\!\!\!\!\!\!\!\!\!\!\!\!\!\!\!\!+ P_B^2 \left({\rm B}^{M}_0[M^{2},0,M^{2}] -{\rm B}^{lM}_0[u,m^2,M^{2}] \right) \frac{}{}\right] \ .
\end{eqnarray}

\section{Coefficients of the IB integrals}

In this Appendix we provide the expressions for the non-vanishing coefficients that enter the definition of the  model-independent real-photon corrections in Eq. (\ref{bremss1}). They are different from the analogous coefficientes semileptonic kaon decays \cite{Ginsberg}.

\begin{eqnarray}
C_{-1,0}&=&\frac{16}{m_V^2} \left[2 a_1 E_V M (a+a_2) +M^2 \left(m_V^2 v^2\right. \right. \nonumber \\
&& \left. \left. -(a+a_2)^2 \vec{P}_V^2\right)-a_1^2 \right]
\end{eqnarray}

\begin{eqnarray}
C_{0,-1} &=& \frac{32}{m_V^2} \left[-a_1 E_V M (a+3 a_2)+a_1 m_V^2 (a+a_2) \right. \nonumber \\
&& +M \left(2 a_2 M (a+a_2) \vec{P}_V^2  +m_V^2 v^2 (E_V-M)\right) \nonumber \\
&&\left. +a_1^2\right]
\end{eqnarray}

\begin{eqnarray}
C_{0,0} &=& \frac{16 M}{m_V^2} \left[2 a_1 \left(-a (2 E+E_V-M) \left(E_V M-m_V^2\right) \right. \right.  \nonumber \\
&& +a_2 E_V M (-6 E-3 E_V+M) \nonumber \\
&& \left. +a_2 m_V^2 (2 E+E_V+M) +2 m_V^2 v (M-E_V)\right) \nonumber \\
&& +M \left(4 a_2 M \vec{P}_V^2 ((a+a_2) (2 E+E_V)-a M) \right. \nonumber \\
&& +m_V^2 v^2 \left(-2 M (2 E+E_V)+2 E_V (2 E+E_V) \right. \nonumber \\
&& \left. \left. \left. +M^2-m_V^2\right)\right)+a_1^2 (4 E+2 E_V-M)\right]
\end{eqnarray}

\begin{eqnarray}
C_{1,1} &=& -\frac{4 E M x}{m_V^2} \left[2 a_1 M \left(a \left(-E E_V M+E m_V^2+E_V m^2\right) \right. \right. \nonumber \\
&& +a_2 E_V \left(-3 E M+m^2+2 M^2\right) \nonumber \\
&& \left. +a_2 m_V^2 (E-2 M)-2 m_V^2 v (E+E_V-M)\right) \nonumber \\
&& +M^2 \left(m_V^2 v^2 \left(-2 M (E+E_V)+2 E_V (E+E_V) \right. \right. \nonumber \\
&& \left. +(m-m_V) (m+m_V)+M^2\right)  \nonumber \\
&& \left. -\vec{P}_V^2 \left(-4 a_2 E M (a+a_2) +m^2 (a+a_2)^2+4 a_2^2 M^2\right)\right) \nonumber \\
&& \left. -a_1^2 \left(-2 E M+m^2+M^2-2 m_V^2\right)\right]
\end{eqnarray}

\begin{eqnarray}
C_{-1,1} &=& -\frac{16 M }{m_V^2} \left[a_1 M \left(a \left(E_V (M-2 (E+E_V))+m_V^2\right) \right. \right. \nonumber \\
&&\left. +a_2 \left(m_V^2-E_V (2 (E+E_V)+M)\right)+2 m_V^2 v\right) \nonumber \\
&& +M^2 \left((a+a_2) \vec{P}_V^2 \right. \nonumber \\ 
&& ~ \times (a (E+E_V-M)+a_2 (E+E_V+M)) \nonumber \\
&& \left. \left. -E m_V^2 v^2\right)+a_1^2 (E+E_V)\right]
\end{eqnarray}

\begin{eqnarray}
C_{1,-1} &=& \frac{16}{m_V^2} \left[a_1 \left(a m^2 \left(E_V M-m_V^2\right) \right. \right. \nonumber \\
&& -a_2 m_V^2 \left(4 M (3 E+E_V-M)+m^2\right) \nonumber \\
&& +a_2 E_V M \left(4 M (3 E+E_V-M)+3 m^2\right) \nonumber \\
&& \left. -2 m_V^2 v \left(-2 E_V M+M^2+m_V^2\right)\right) \nonumber \\
&& +M \left(m_V^2 v^2 \left(E_V \left(3 M (M-2 E)-m^2+m_V^2\right) \right. \right. \nonumber \\
&& \left. +(3 E-M) \left(M^2+m_V^2\right)-2 E_V^2 M+m^2 M\right) \nonumber \\
&& -2 a_2 M \vec{P}_V^2 \nonumber \\ 
&& ~ \times \left. \left(a m^2+a_2 \left(2 M (3 E+E_V-M)+m^2\right)\right)\right) \nonumber \\
&& \left. -a_1^2 \left(M (3 E+E_V-M)+m^2\right)\right]
\end{eqnarray}

\begin{eqnarray}
C_{2,-1} &=& \frac{16 m^2}{m_V^2} \left[ a_1 \left(a m^2 \left(E_V M-m_V^2\right) \right. \right. \nonumber \\
&& -a_2 m_V^2 \left(4 M (-2 E-E_V+M)+m^2\right) \nonumber \\
&& +a_2 E_V M \left(4 M (-2 E-E_V+M)+3 m^2\right) \nonumber \\
&& \left. +2 m_V^2 v \left(-2 E_V M+M^2+m_V^2\right)\right) \nonumber \\
&& +M \left(m_V^2 v^2 \left(-E_V \left(-4 E M+m^2+3 M^2+m_V^2\right) \right. \right. \nonumber \\
&& \left. -(2 E-M) \left(M^2+m_V^2\right)+2 E_V^2 M+m^2 M\right) \nonumber \\
&& -2 a_2 M \vec{P}_V^2 \nonumber \\ 
&& ~ \times \left. \left(m^2 (a+a_2)-2 a_2 M (2 E+E_V)+2 a_2 M^2\right)\right) \nonumber \\
&& \left. -a_1^2 \left(M (-2 E-E_V+M)+m^2\right)\right]
\end{eqnarray}

\begin{eqnarray}
C_{2,-2} &=& \frac{16 m^2}{m_V^2} \left[a_1^2+4 a_1 a_2 \left(m_V^2-E_V M\right) \right. +4 a_2^2 M^2 \vec{P}_V^2 \nonumber \\
&& \left. -m_V^2 v^2 \left(-2 E_V M+M^2+m_V^2\right)\right]
\end{eqnarray}

\begin{eqnarray}
C_{0,2} &=& \frac{4 M^2 x}{m_V^2} \left[ 2 a_1 M \left(a \left(-E E_V M+E m_V^2+E_V m^2\right) \right. \right. \nonumber \\
&& +a_2 E_V \left(-3 E M+m^2+2 M^2\right) \nonumber \\
&&\left. +a_2 m_V^2 (E-2 M)-2 m_V^2 v (E+E_V-M)\right) \nonumber \\
&& +M^2 \left(m_V^2 v^2 \left(-2 M (E+E_V)+2 E_V (E+E_V) \right. \right. \nonumber \\
&& \left. +(m-m_V) (m+m_V)+M^2\right) \nonumber \\
&& -\vec{P}_V^2 \left(-4 a_2 E M (a+a_2) \right. \nonumber \\
&& \left. \left. +m^2 (a+a_2)^2 +4 a_2^2 M^2\right)\right) \nonumber \\
&& \left. -a_1^2 \left(-2 E M+m^2+M^2-2 m_V^2\right)\right]
\end{eqnarray}

\begin{eqnarray}
C_{-1,2} &=& -\frac{8 M^2}{m_V^2} \left[ M^2 \left(a^2 \vec{P}_V^2  (x-m^2) \right. \right. \nonumber \\
&& +2 a a_2 \vec{P}_V^2 \left(2 M (E+E_V-M)-m^2+x\right) \nonumber \\
&& +a_2^2 \vec{P}_V^2 \left(4 M (E+E_V)-m^2+x\right) \nonumber \\
&& +m_V^2 v^2 \left(-2 M (E+E_V)+2 E E_V \right. \nonumber \\
&& \left. \left. +m^2+M^2+m_V^2-x\right)\right) \nonumber \\
&& +2 a_1 M \left(a \left(E_V \left(-E M+m^2+M^2+m_V^2-x\right) \right. \right. \nonumber \\
&& \left. +m_V^2 (E_{\ell}-M)-E_V^2 M\right) \nonumber \\
&& +a_2 \left(E_V \left(-3 E M+m^2+M^2+m_V^2-x\right) \right. \nonumber \\
&& \left. \left. +m_V^2 (E+M)-3 E_V^2 M\right)+2 E m_V^2 v\right) \nonumber \\
&& -a_1^2\left(M (M-2 (E+E_V)) \right. \nonumber \\
&& \left. \left. +m^2+2 m_V^2-x\right)\right]
\end{eqnarray}

\begin{eqnarray}
C_{1,-2} &=& -\frac{16}{m_V^2} \left[ a_1^2+4 a_1 a_2 \left(m_V^2-E_V M\right)+4 a_2^2 M^2 \vec{P}_V^2 \right. \nonumber \\
&& \left. -m_V^2 v^2 \left(-2 E_V M+M^2+m_V^2\right)\right]
\end{eqnarray}

\begin{eqnarray}
C_{0,1} &=& -\frac{8 M}{m_V^2} \left[ M^2 \left(a^2 \vec{P}_V^2 \left(m^2 (3 E+E_V-M)-E x\right) \right. \right. \nonumber \\
&& -2 a a_2 \vec{P}_V^2 \left(6 E^2 M \right. \nonumber \\
&& +E \left(6 M (E_V-M)-3 m^2+x\right) \nonumber \\
&& \left. -m^2 (E_V+M)+M x\right) \nonumber \\
&& -a_2^2 \vec{P}_V^2 \left(12 E^2 M \right. \nonumber \\ 
&& +E \left(12 E_V M-3 m^2+8 M^2+x\right) \nonumber \\
&& \left. -E_V m^2+4 E_V M^2-3 m^2 M-4 M^3+2 M x\right) \nonumber \\
&& +m_V^2 v^2 \left(6 E^2 (M-E_V) \right. \nonumber \\
&& +E \left(-4 E_V^2+2 E_V M-3 m^2-M^2+3 m_V^2+x\right) \nonumber \\
&& -(E_V-M) \nonumber \\ 
&&  ~~\left. \left. \times \left(2 E_V^2+2 E_V M-m^2-M^2-3 m_V^2+x\right)\right)\right) \nonumber \\
&& -a_1 M \left(a \left(6 E^2 \left(m_V^2-E_V M\right) \right. \right. \nonumber \\
&& +E \left(-6 E_V^2 M+6 E_V \left(m^2+M^2+m_V^2\right) \right. \nonumber \\
&& \left. -2 E_V x-6 M m_V^2\right) \nonumber \\
&& \left. +2 m^2 \vec{P}_V^2+x \left(m_V^2-E_V M\right)\right) \nonumber \\
&& +a_2 \left(6 E^2 \left(m_V^2-3 E_V M\right) \right. \nonumber \\
&& -2 E \left(9 E_V^2 M+E_V \left(-3 m^2+M^2-3 m_V^2+x\right) \right. \nonumber \\
&& \left. -7 M m_V^2\right)+2 E_V^2 \left(m^2-2 M^2\right) \nonumber \\
&& +E_V M \left(4 \left(m^2+M^2+m_V^2\right)-3 x\right) \nonumber \\
&& \left. +m_V^2 \left(-2 m^2-4 M^2+x\right)\right) \nonumber \\
&& \left. +2 m_V^2 v \left(2 \left(E^2+2 E (M-E_V)-\vec{P}_V^2\right)-x\right)\right) \nonumber \\
&& +a_1^2 \left(-6 E^2 M+E \left(-6 E_V M+3 m^2+M^2 \right. \right. \nonumber \\
&& \left. +2 m_V^2-x\right)+E_V \left(m^2-M^2-2 m_V^2\right) \nonumber \\
&& \left. \left. +M \left(m^2+M^2+2 m_V^2-x\right)\right)\right] 
\end{eqnarray}

\begin{eqnarray}
C_{1,0} &=& \frac{4}{m_V^2} \left[-M^2 \right. \nonumber \\
&& ~ \times \left(-a^2 m^2 \vec{P}_V^2 \left(2 E_V M+3 m^2-M^2  -m_V^2\right) \right. \nonumber \\
&& +2 a a_2 \vec{P}_V^2 \left(m^2 \left(M (-2 E+2 E_V-3 M)+m_V^2\right) \right. \nonumber \\
&& \left. -2 E M \left(-2 E_V M+M^2+m_V^2-x\right)-3 m^4\right) \nonumber \\
&& +a_2^2 \vec{P}_V^2 \left(4 M \left(16 E^2 M+E \left(10 E_V M-9 M^2 \right. \right. \right. \nonumber \\
&& \left. \left. -m_V^2+x\right)+M \left(-2 E_V M+M^2+m_V^2\right)\right) \nonumber \\
&& \left. +m^2 \left(M (-4 E+6 E_V-3 M)+m_V^2\right)-3 m^4\right) \nonumber \\
&& +m_V^2 v^2 \left(16 E^2 \left(2 E_V M-M^2-m_V^2\right) \right. \nonumber \\
&& +2 E (E_V-M) \left(10 E_V M-m^2-5 \left(M^2+m_V^2\right) \right. \nonumber \\
&& \left. +x\right) +4 E_V^3 M+2 E_V^2 \left(m^2-3 M^2-m_V^2\right) \nonumber \\
&& +4 E_V M \left(M^2-m^2\right) \nonumber \\
&& \left. \left. +3 m^4+2 m^2 M^2-M^4+m_V^4\right)\right) \nonumber \\
&& +2 a_1 M \left(a \left(2 E E_V^2 M^2 \right. \right. \nonumber \\
&& -E_V \left(m^2 \left(M (E+M)+m_V^2\right) \right. \nonumber \\
&& \left. +E M \left(M^2+3 m_V^2-x\right)+3 m^4\right) \nonumber \\
&& \left. +m_V^2 \left(E \left(m^2+M^2+m_V^2-x\right)+2 m^2 M\right)\right) \nonumber \\
&& +a_2 \left(-E_V \left(-M \left(32 E^2 M-19 E M^2-21 E m_V^2 \right. \right. \right. \nonumber \\
&& \left. +3 E x+2 M^3+6 M m_V^2\right) \nonumber \\
&& \left. +m^2 \left(3 M (E+M)+m_V^2\right) +3 m^4\right) \nonumber \\
&& +m_V^2 \left(-32 E^2 M \right. \nonumber \\
&& \left. +E \left(m^2+17 M^2+m_V^2-x\right)-2 M \left(M^2+m_V^2\right)\right) \nonumber \\
&& \left. +2 E_V^2 M \left(M (11 E-2 M)+2 m^2\right)\right) \nonumber \\
&& +2 m_V^2 v \left(E \left(10 E_V M-m^2-5 \left(M^2+m_V^2\right)+x\right) \right. \nonumber \\
&& \left. \left. +(E_V-M) \left(2 E_V M+m^2-M^2-m_V^2\right)\right)\right) \nonumber \\
&& +a_1^2 \left(-16 E^2 M^2+m^2 \left(2 M (E-E_V+M)+m_V^2\right) \right. \nonumber \\
&& +2 E M \left(-6 E_V M+5 M^2+m_V^2-x\right)+3 m^4 \nonumber \\
&& \left. \left. -\left(M^2-2 m_V^2\right) \left(-2 E_V M+M^2+m_V^2\right) \right)\right]
\end{eqnarray}
{In the above expressions we have defined $\vec{P}_V^2=E_V^2-m_V^2$ and have introduced the following reduced form factors: \ $v=2 V/(M + m_{\rho}),\ a_1= (M+m_{\rho}) A_{1}, \ a_2= A_{2}/(M + m_{\rho})$ and $a=2 m_{\rho} A/q^{2} $.




\end{document}